\documentclass{PoS}
\usepackage{amsmath,bm}

\title{Excited-state contamination in nucleon correlators from chiral perturbation theory}

\ShortTitle{Excited-state contamination $\in$ nucleon correlators}

\author{\speaker{Brian C.~Tiburzi}
\thanks{Work supported in part by a joint The City College of New York--RIKEN/BNL Research Center fellowship, 
an award from the Professional Staff Congress of The CUNY, 
and by the U.S.~National Science Foundation, 
under Grant No.~PHY$15$-$15738$. 
}\\
        Department of Physics,
        The City College of New York,  
        New York, NY, USA\\
        Graduate School and University Center,
        The City University of New York,
        New York, NY, USA\\
        RIKEN BNL Research Center, 
        Brookhaven National Laboratory, 
        Upton, NY, USA\\
        E-mail: \email{btiburzi@ccny.cuny.edu}}

\abstract{Techniques to compute hadron properties from lattice QCD rely upon the limit of long time separation. 
For baryons, 
the signal-to-noise problem often restricts one to time separations that are not ideally long, 
and for which couplings to excited states can obstruct the isolation of  ground-state baryon properties. 
We consider excited-state contamination in nucleon two- and three-point functions. 
Using chiral perturbation theory, 
we determine couplings to pion-nucleon and pion-delta excited states. 
In two-point functions, 
these contributions are small, 
in accordance with general properties of the spectral weights on a torus. 
For the axial-current correlation function in the nucleon, 
the sign of excited-state contributions suggests overestimation of the nucleon axial charge. 
Thus contamination from pion-nucleon excited states will not likely explain the trend in lattice QCD data.}

\FullConference{The 8th International Workshop on Chiral Dynamics, CD2015\\
		29 June 2015 - 03 July 2015\\
		Pisa, Italy}

\begin{document}

\section{Motivation and Overview}

The motivation to study excited-state contamination in nucleon correlation functions using chiral perturbation theory comes directly from lattice QCD. 
As a result, 
our primary concern is with the Euclidean time dependence of correlation functions, 
such as two-point and three-point functions that are relevant to extract the nucleon mass and current matrix elements using standard lattice QCD techniques. 
For a rundown of such techniques relevant to baryon physics, 
see%
~\cite{Lin:2015dga}. 
In the case of the nucleon two-point function, 
large Euclidean time separation is used as a filter for the ground-state nucleon. 
From the zero three-momentum projected correlator, 
$G(\tau)$, 
one typically looks at the effective mass
\begin{equation}
M_{\text{eff}}(\tau)
\equiv
- \log \frac{G(\tau + a)}{G(\tau)}
\overset{\tau \to \infty}{=}
M_N + |Z'|^2 e^{- \Delta E \,  \tau} + \cdots
\label{eq:effmass}
,\end{equation}
where 
$a$ 
is the lattice spacing, 
$M_N$
is the nucleon mass, 
and $\Delta E$
is the energy splitting with the nearest excited state, 
$\Delta E = E' - M_N$, 
with 
$|Z'|^2$ as the ratio of the excited-state to ground-state coupling of the chosen lattice nucleon interpolating operator. 
A flat effective mass indicates that the ground state saturates the correlation function, 
and allows for a reliable extraction of the nucleon mass.  
To reduce uncertainty surrounding saturation of the ground state, 
one should go to suitably long Euclidean time separations, 
however, 
lattice QCD correlation functions are determined stochastically. 
For baryon correlation functions, 
it is well known%
~\cite{Lepage:1989hd}
that the signal in 
Eq.~\eqref{eq:effmass} 
degrades exponentially compared with the statistical noise.  
Without an increase in statistics, 
one can attempt to minimize 
$|Z'|^2$
through the construction of better nucleon interpolating operators. 
Such operators couple less strongly to excited states; 
and, 
to this end, 
we investigate the nature of pion-nucleon excited states using chiral perturbation theory. 
One should note that the signal-to-noise problem is more restrictive in the case of three-point functions, 
due to the necessity of 
\emph{two} 
Euclidean time separations that should be ideally long.

Here we address the pion-nucleon (and, to a lesser extent, pion-delta) contributions to nucleon correlation functions. 
The un-amputated correlators are computed as a function of Euclidean time using chiral perturbation theory. 
This is a summary of previous work, namely: the computation of excited-state contamination using a continuum of excited states was pursued in%
~\cite{Tiburzi:2009zp}, 
while this computation was recently revisited for the case of discrete pion-nucleon and pion-delta states available on a periodic lattice%
~\cite{Tiburzi:2015tta}. 
We detail the computation of the nucleon two-point correlation function, 
because it contains universal couplings to pion-nucleon states. 
We also investigate the excited-state contamination present in the nucleon three-point function of the isovector 
axial-vector current.  
The forward matrix element of this current is parameterized by the nucleon axial charge, 
$g_A$. 
We find the sign of excited-state contamination in 
$g_A$
is inconsistent with underestimation of the axial charge. 
An outlook to future work is given in the summary.

\section{Nucleon Operators}

\begin{figure}
\begin{center}
\includegraphics[width=0.33\textwidth]{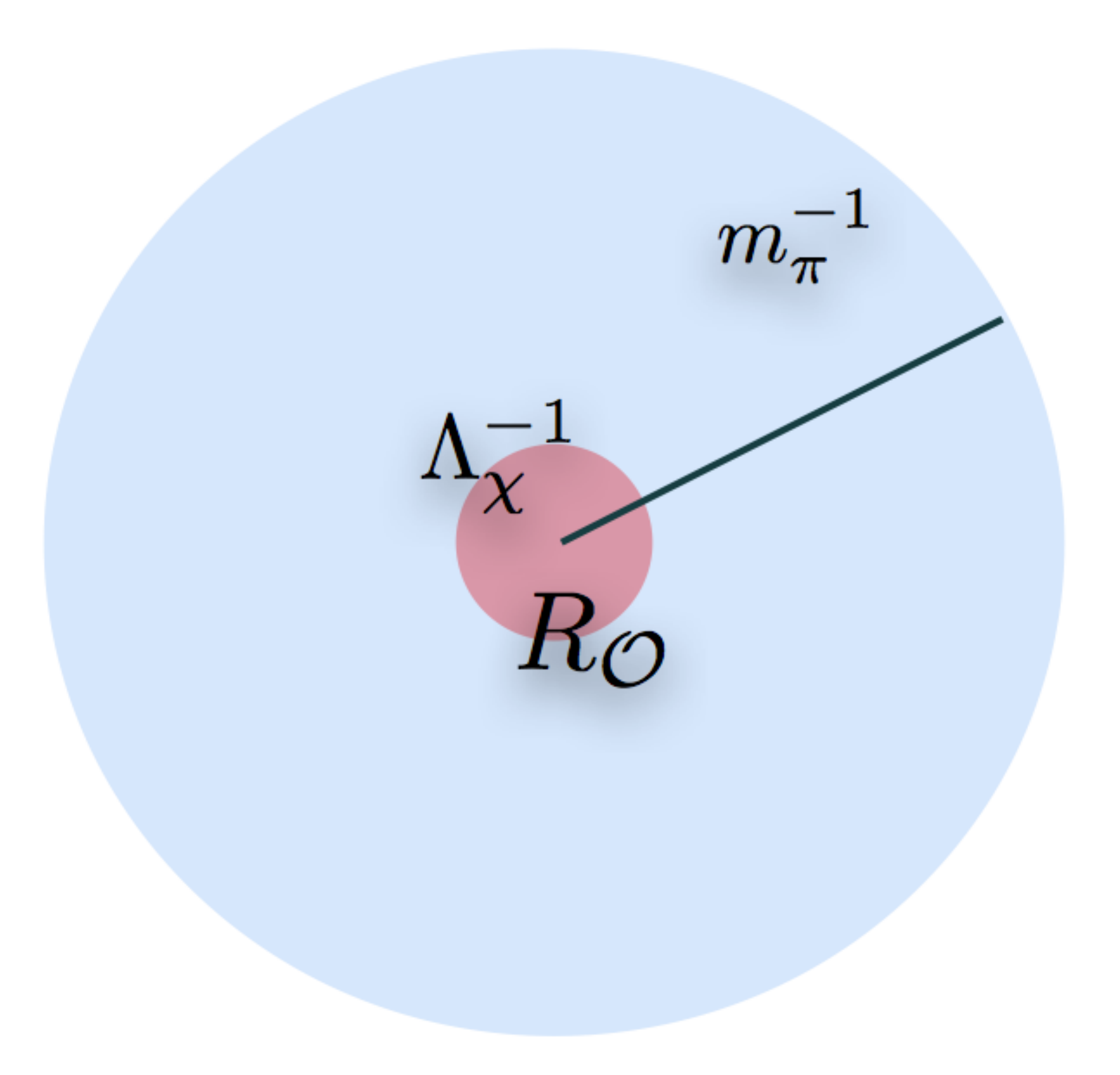}
\caption{
Graphical depiction of the nucleon operator in chiral perturbation theory, 
as well as a lattice QCD interpolating operator for the nucleon. 
In the former,
the size of the short-distance core is set by 
$\Lambda_\chi^{-1}$, 
while in the latter, 
the operator smearing radius 
$R_\mathcal{O}$
sets the size. 
}
\label{f:rest}
\end{center}
\end{figure}

To begin, 
we note that within chiral perturbation theory the nucleon is described by a local operator, 
$N(x)$. 
Of course locality is only defined relative to some scale, 
and this scale is set by the inverse of the chiral symmetry breaking scale
$\Lambda_\chi = 2 \sqrt{2} \pi f$, 
where 
$f$
is the pion decay constant
and
$f = 132 \, \texttt{MeV}$. 
The long-distance nature of the pion means that observables in chiral perturbation theory 
are computed as long-range pion contributions plus short-range physics, 
see Fig.~\ref{f:rest}. 
In the effective field theory, 
we know that the nucleon operator transforms as a doublet under the unbroken isospin symmetry, 
$SU(2)_V$. 
It is, 
however, 
unknown to which chiral multiplet(s) the nucleon belongs in the chiral limit. 
In fact, 
the transformation property of the nucleon field under 
$SU(2)_L \times SU(2)_R$
is not an ingredient of chiral perturbation theory. 
It is both customary and convenient to have the nucleon field transform as
$N (x) \to U\Big( L, R, \xi(x) \Big) N(x)$
under 
$L,R \in SU(2)_L \times SU(2)_R$, 
where 
$\xi$
is the square-root of the coset field,
$\xi = \sqrt{\Sigma}$, 
and transforms as 
$\xi \to U \xi R^\dagger = L \xi U^\dagger$.
This freedom stems from the ambiguity in resolving a nucleon and any number of pion states in the chiral limit. 
Using this simple transformation, 
the pion-nucleon (and pion-delta) interactions are then easily constrained by the pattern of spontaneous and explicit chiral symmetry breaking of QCD.

%
%
%
%

In lattice QCD, the situation is actually quite analogous. 
One studies properties of the nucleon using a lattice interpolating operator, 
$\mathcal{O}_N (x)$.  
To overlap with the nucleon, 
this operator must have the correct isospin quantum numbers. 
Unlike chiral perturbation theory, 
however,
lattice interpolating operators belong to a given chiral multiplet. 
The chiral properties of baryon operators have been studied completely%
~\cite{Nagata:2008zzc}.
We will use the most common of such operators
\begin{equation}
\mathcal{O}_N
\sim 
q \left( q^T C \gamma_5 \tau^2 q \right)
,\end{equation}
where 
$\vec{\tau}$
are the isospin matrices, 
and 
$q$ 
is an isodoublet of quark fields, 
$q = \begin{pmatrix} u \\ d \end{pmatrix}$. 
This operator transforms as
$(\bm{2}_L, \bm{1}_R ) \oplus (\bm{1}_L, \bm{2}_R )$
under chiral transformations. 
Typically nucleon operators are chosen to maximize overlap with the ground-state nucleon, 
which is characterized by the overlap factor
\begin{equation}
Z_\mathcal{O}
=
\langle 0  | \mathcal{O}_N | N \rangle
.\end{equation}
There are also excited-state couplings of the operator, 
for example
$Z_{\pi N} = \langle 0 | \mathcal{O}_N | (\pi N)_p \rangle$, 
where the 
$p$
denotes a relative 
$p$-wave. 
One way to maximize overlap with the ground state is to smear the operator over some size 
$R_\mathcal{O}$. 
It turns out that the relative contribution of the pion-nucleon excited states compared to the ground state, 
$|Z_{\pi N}|^2 / |Z_N|^2$,
is constrained by chiral dynamics provided the operator smearing radius is small compared to the pion Compton wavelength, 
$R_\mathcal{O} \ll m^{-1}$%
~\cite{Bar:2015zwa}. 
A similar idea was put forth for gradient flow observables by considering the flow time as a scale in chiral perturbation theory, 
see%
~\cite{Bar:2013ora}.

The above discussion implies the procedure to compute the chiral contamination in nucleon correlation functions. 
For a given lattice interpolating operator, 
$\mathcal{O}_N(x)$, 
we assume the hierarchy of scales
$R_\mathcal{O} \sim M_N^{-1} \sim \Lambda_\chi^{-1} \ll m_\pi^{-1}$, 
so that the long-range physics remains dominated by pions. 
We then employ chiral perturbation theory for the lattice operator, 
by mapping the lattice interpolating field into effective field theory operators, 
$N_\mathcal{O} = N_{\mathcal{O}} ( N, \xi )$. 
This mapping is also required in the case of chiral perturbation theory for moments of the nucleon distribution amplitudes%
~\cite{Wein:2011ix}. 
In that case, 
of course,
the nucleon operators are strictly local.%
\footnote{
Chiral perturbation theory calculations of excited-state contamination have been performed for two-point correlation functions of local axial-vector and pseudoscalar bilinear operators in vacuum%
~\cite{Bar:2012ce}.
} 
We then use the effective field theory nucleon operators to compute correlation functions. 
As an example, 
nucleon chiral doublet operators take the form
\begin{eqnarray}
N_{( \bm{2}_L, \bm{1}_R)}
&\sim&
\frac{1}{2} \xi N + \mathcal{O} (M_N^{-1})
,\quad \text{and} \quad
N_{( \bm{1}_L, \bm{2}_R)}
\sim
\frac{1}{2} \xi^\dagger N + \mathcal{O} (M_N^{-1})
.\end{eqnarray}
This leads to the follow heavy nucleon effective field theory operator for the lattice interpolator 
\begin{equation}
N_{( \bm{2}_L, \bm{1}_R) \oplus ( \bm{1}_L, \bm{2}_R)}
=
\frac{1}{2} Z_\mathcal{O} ( \xi + \xi^\dagger) N
=
Z_\mathcal{O}
\left( 1 - \frac{\pi^2}{2 f^2} + \cdots \right) N
.\end{equation}
From this expression, 
we can see that the relative coupling of pions to the nucleon is fixed for the given lattice operator because of that operator's chiral transformation properties. 
For a given nucleon correlation function, 
there will be pion loop diagrams that contribute in addition to the pions generated from the chiral structure of the lattice operator.

\section{Chiral Contamination in the Nucleon Two-Point Function}

\begin{figure}
\begin{center}
\includegraphics[width=0.5\textwidth]{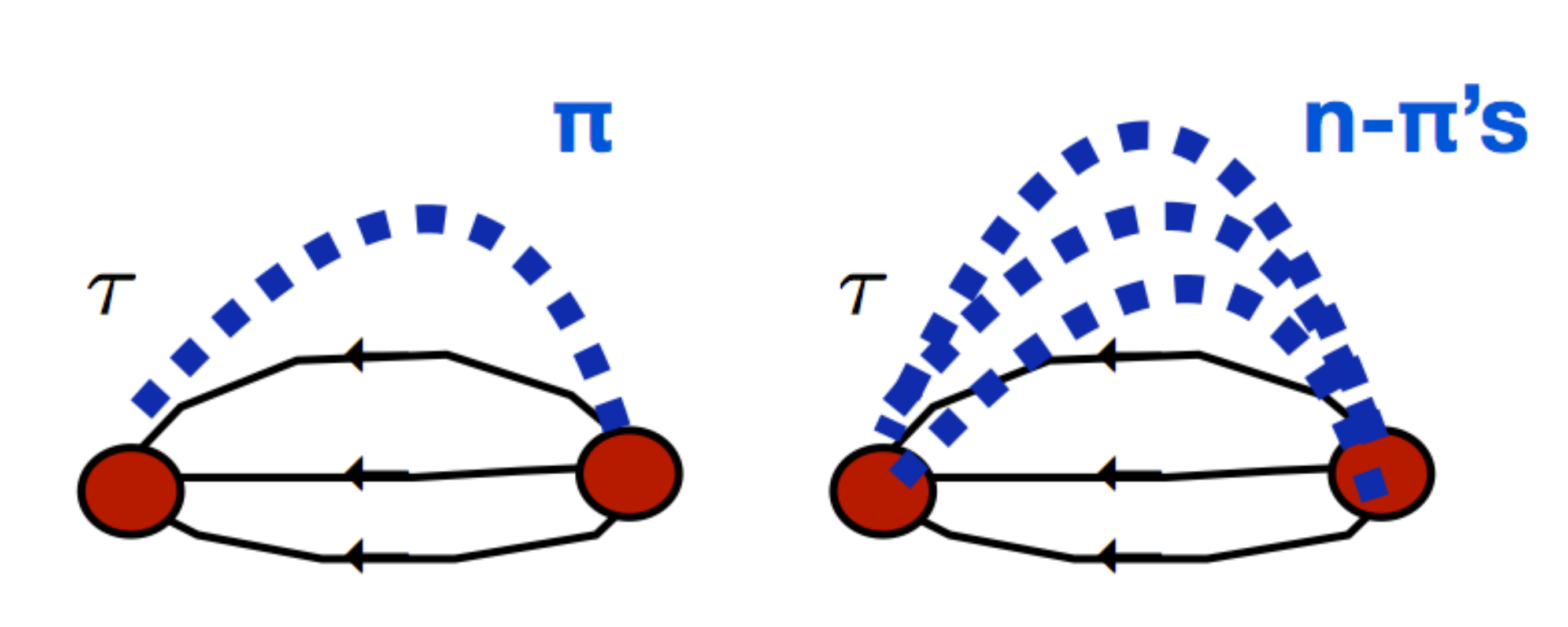}
\caption{
Feynman diagrams depicting excited-state contamination in the nucleon two-point function. 
A single pion shown on the left, while an intermediate state with multiple pions is shown on the right. 
}
\label{f:pions}
\end{center}
\end{figure}

Now we turn our attention to the determination of the nucleon two-point function using chiral perturbation theory. 
This two-point function has a matrix element definition
\begin{equation}
G(\tau)
=
\sum_{\vec{x}}
\langle 0 | \mathcal{O}_N(\vec{x}, \tau) \mathcal{O}^\dagger_N (\vec{0}, 0) | 0 \rangle
,\end{equation}
where the sum over all lattice sites
$\vec{x}$
projects the correlation function onto vanishing three momentum. 
Quite generally the two-point function has a spectral representation with positive spectral weights representing the probability to find excited states in the correlator.  
For our purposes, 
we write the spectral representation in the form 
\begin{equation}
G(\tau)
= 
|Z_\mathcal{O}|^2
e^{ - M_N \tau}
\left[
1
+ 
\sum_{n\neq0} 
|Z_n|^2 
e^{ - \Delta E_n \tau}
\right]
\label{eq:2}
,\end{equation}
where the excited-state couplings
$|Z_n|^2$
are defined relative to the ground-state nucleon, 
as is the energy 
$\Delta E_n$, 
which has the form 
$\Delta E_n = E_n - E_0$, 
where 
$E_0 = M_N$
is the ground-state energy, 
which is just the nucleon mass.  
Chiral perturbation theory makes a prediction for the relative contribution, 
$|Z_n|^2$,
from the various low-lying excited states. 
Our primary concern is with intermediate pion-nucleon states, 
see Fig.~\ref{f:pions}.

%
%
%
%

The spectral density has previously been determined in chiral perturbation theory%
~\cite{Tiburzi:2009zp}.
In that work, 
a continuum of pion-nucleon states was employed, 
for which one has the relation
$\sum_{n} |Z_n|^2 \to \int_{m_\pi}^\infty dE \, \rho_{\pi N} (E)$. 
The spectral density has the form
\begin{equation}
\rho_{\pi N} (E)
=
\frac{6 g_A^2}{( 4 \pi f)^2}
\frac{|\vec{k}_\pi|^{3}}{\vec{k}_\pi^2 + m_\pi^2}
,\end{equation}
which has a number of features. 
First is the non-analyticity at threshold, 
such non-analyticity can only arise from a continuum of states. 
Second is the behavior near threshold, 
$|\vec{k}_\pi|^{3}$, 
which arises because of the two-body phase space
$\sim | \vec{k}_\pi |$, 
and the requirement that the nucleon and pion be in a relative 
$p$-wave, 
$\sim \vec{k}_\pi^2$.
By contrast, 
on a periodic spatial lattice of length
$L$
in each direction, 
the allowed pion momentum modes are restricted to 
$\vec{k}_\pi = \frac{2 \pi}{L} \vec{n}$. 
For such modes, 
we have the weights%
~\cite{Tiburzi:2015tta}
\begin{equation}
|Z_{\pi N}|^2
=
\frac{3 g_A^2}{4 f^2 L^2} 
\frac{\vec{k}_\pi^2}{[\vec{k}_\pi^2 + m_\pi^2]^{3/2}}
\label{eq:specweigh}
,\end{equation}
which are analytic at threshold; because, there, we no longer have an accumulation of states.
Thus at threshold, 
the behavior of the two-body relative weights is determined solely by the requirements of angular momentum. 
It should be noted that the two-point function of the nucleon has been determined using 
relativistic baryon chiral perturbation theory%
~\cite{Bar:2015zwa}. 
Expanding these results in 
$m_\pi / M_N \sim 0.15$,
they reduce to those given in Eq.~\ref{eq:specweigh}.

%
%
%
\begin{figure}
\centering
\includegraphics[width=0.5\textwidth]{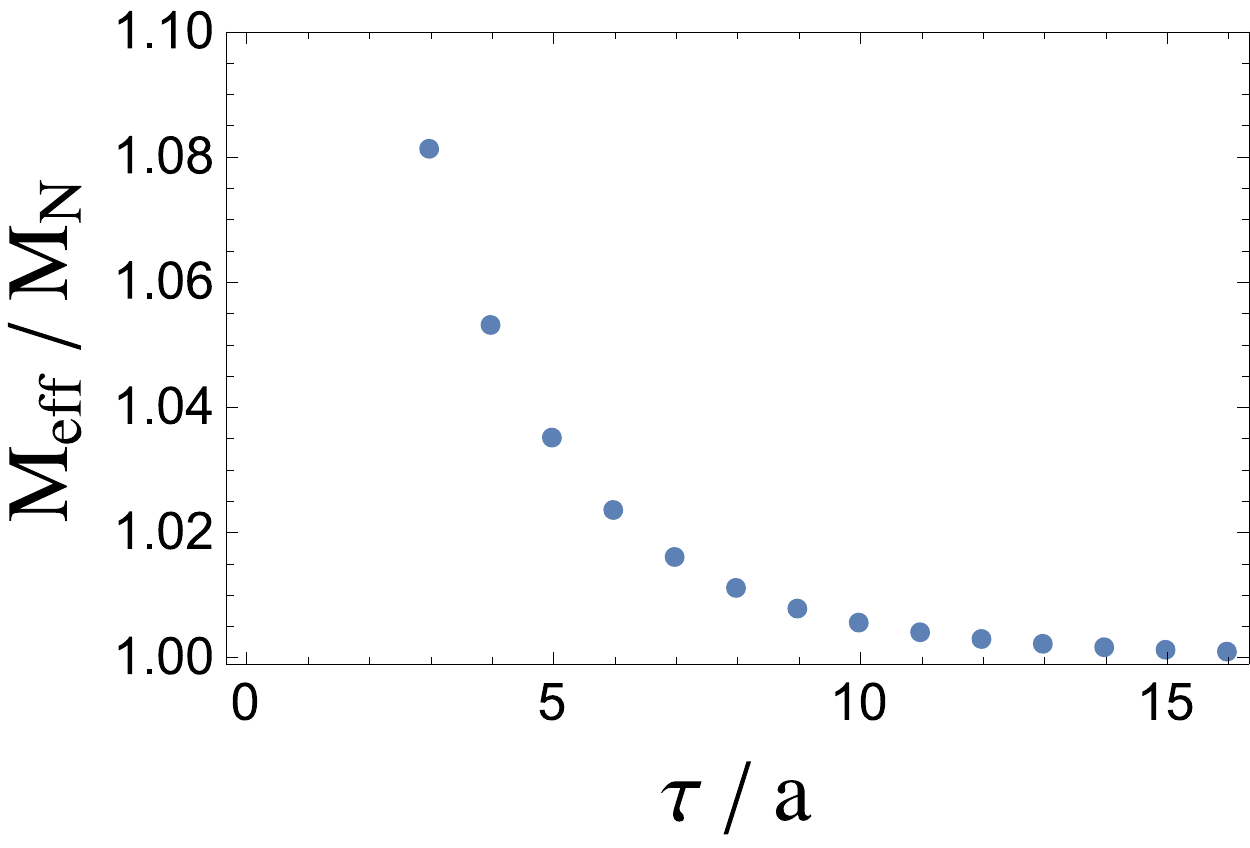}%
\caption{
Effective mass of the nucleon two-point function computed with chiral perturbation theory.
Pion-nucleon excited-state contamination is shown at the few-percent level for typical values of lattice parameters. 
For times such that
$\tau / a > 5$, 
the correlator is saturated by pion-nucleon states with momentum mode numbers satisfying
$\sqrt{\vec{n}^2} < 4$. 
}
\label{f:effm}
\end{figure}
%
%
%

The effective mass of the two-point correlation function in Eq.~\ref{eq:2} 
is plotted in Fig.~\ref{f:effm}. 
For the relative spectral weights, 
we use those derived from chiral pertubation theory, 
Eq.~\ref{eq:specweigh}. 
To make the plot, 
we use the physical value of the pion mass, 
and choose a lattice spacing of 
$a = 0.1 \, \texttt{fm}$
on a lattice of size
$L = 48 a$. 
From the plot, 
we see that pion-nucleon excited states do not contribute substantially to the two-point correlation function. 
In fact, 
the suppression of pion-nucleon states, 
and more generally (pion)${}^n$-nucleon states 
(see Fig.~\ref{f:pions})
can be argued without any reference to chiral perturbation theory. 
The reasoning stems from simple phase-space considerations near threshold. 
For a continuum of states, 
the spectral density near threshold is determined entirely by the available phase space, 
and the orbital angular momentum 
$\ell$
of the multi-particle state. 
As a result,
we have%
\footnote{
Here we define $n$ to be the number of pions, 
whereas in%
~\cite{Tiburzi:2015tta}, 
we use the integer
$N$ to denote the number of particles in the intermediate state, 
which is thus
$n+1$
for the present case.
}
\begin{equation}
\rho_{(\pi)^n N} (E)
\propto 
\sqrt{E - E_\text{Th}}^{3 n +  2 \ell -2}
.\end{equation}
Indeed there is an analogous result for the spectral weights of such 
$n$-pion states available on a spatial torus. 
The suppression of such multi-particle states is comparatively less than for a continuum of states,
as the weights take the form%
\footnote{
One interesting feature is that there is no threshold suppression from phase space of 
$\pi N$-states in a relative $s$-wave on a torus. 
}
\begin{equation}
|Z_{(\pi)^n N}|^2
\propto
\sqrt{E - E_{\text{Th}}}^{n + 2 \ell - 1}
.\end{equation}
Chiral perturbation theory implies that these multi-pion states are further suppressed by factors of 
$(g^2_A / f L)^{2n}$.

Phase space arguments near threshold require non-relativistic pion states, 
i.e.~we must satisfy 
$E - E_{\text{Th}} = \frac{\vec{k}_\pi^2}{ 2 m_\pi} + \cdots \ll 1$, 
which translates into the condition
$m_\pi L \gg 1$. 
Suppression of the weights near threshold, 
however,  
can compete with exponential enhancement of excited states in the correlation function. 
This is because enlarging the volume to approach threshold, 
also reduces the energy of the multi-particle state. 
As the gap is lowered,
the exponential factor 
$e^{ - \Delta E \tau}$
increases. 
This may or may not be an issue 
depending on how one approaches the chiral and infinite volume limits. 
In the case of fixed pion mass 
and increasing lattice volume,
there is no issue because 
$m_\pi L$
increases, 
and the decrease in the gap will be negligible compared to 
$m_\pi$. 
On the other hand, 
lowering the pion mass at fixed lattice size is problematic. 
In this case, 
both 
$m_\pi L$ 
and the gap are
decreasing.
Both conditions lead to enhancement of multi-particle states. 
In this situation, 
however, 
one should generally be worried about finite volume effects to begin with.

\section{Axial Current in the Nucleon}

Given the above computation of the two-point function, we now turn our attention to analogous matters in the case of nucleon matrix elements of the iso-vector axial-vector current. 
For simplicity, 
we consider the forward matrix element which is directly proportional to the nucleon axial charge, 
$g_A$.
The un-amputated, 
three-point function of the axial current between nucleon states is defined by
\begin{equation}
\langle 0 | 
\mathcal{O}_N(\tau)
J^+_\mu(t)
\mathcal{O}_N^\dagger(0)
| 0 \rangle 
= 
\left(
\psi^\dagger 2 S_\mu \psi
\right)
\mathcal{G}_A (\tau, t)
,\end{equation}
where the iso-vector axial-vector current is given by
$J^+_\mu = \overline{q} \gamma_\mu \gamma_5  \tau^+ q$ , 
and we have suppressed the projection onto vanishing three-momentum.  
The rest-frame spinors are denoted by 
$\psi$, 
and 
$S_\mu$
is the covariant spin operator. 
The axial current matrix element depends on two Euclidean time separations: 
the source-to-sink separation 
$\tau$, 
and the source-to-current insertion separation, 
$t$.

%
%
\begin{figure}
\begin{center}
\includegraphics[width=0.5\textwidth]{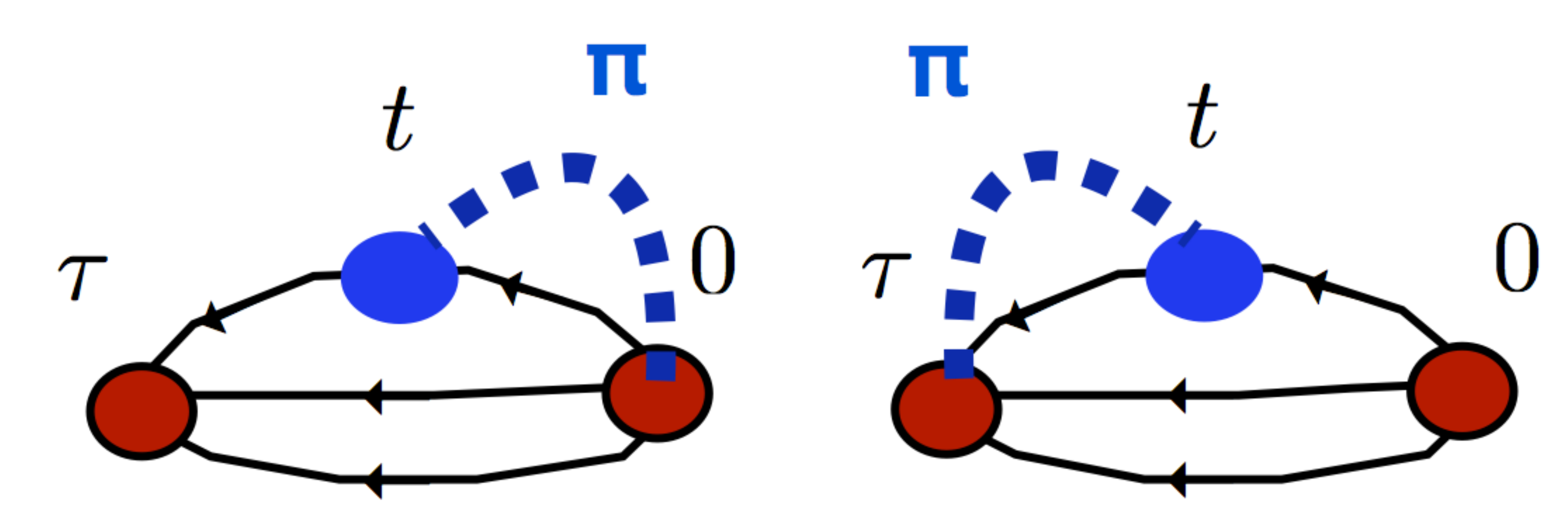}
\caption{
Feynman diagrams depicting the dominant contribution to excited-state contamination in the nucleon three-point function of the axial current.}
\label{f:best}
\end{center}
\end{figure}
%
%

The standard lattice QCD method to compute the axial charge relies on taking the ratio of 
three-point to two-point correlation functions. 
In chiral perturbation theory, 
we define the analogous ratio
\begin{equation}
G_A (\tau, t)
\equiv \mathcal{G}_A (\tau, t) / G(\tau)
.\end{equation}
The time-dependence of this ratio is generally given by an expression of the form
\begin{equation}
G_A(\tau, t)
= 
g_A 
+ 
\sum_{n,m \neq (0,0)}
\mathfrak{g}_{n,m}
\,
e^{ - \Delta E_n (\tau - t)}
e^{ - \Delta E_m (t)}
\label{eq:GA}
,\end{equation}
where 
$\mathfrak{g}_{n,m}$
are the axial transition couplings between the 
$n$-th 
and 
$m$-th excited states. 
And the excited state energies, 
$\Delta E_n$
are measured relative to the nucleon mass.
For Euclidean time separations
$t$
and
$\tau - t$
that are not ideally long, 
the leading excited-state contamination in the three-point function arises from 
$N$-to-$\pi N$ transition couplings, 
as shown in Fig.~\ref{f:best}.

%
%
%
\begin{figure}
\centering
\includegraphics[width=0.7\textwidth]{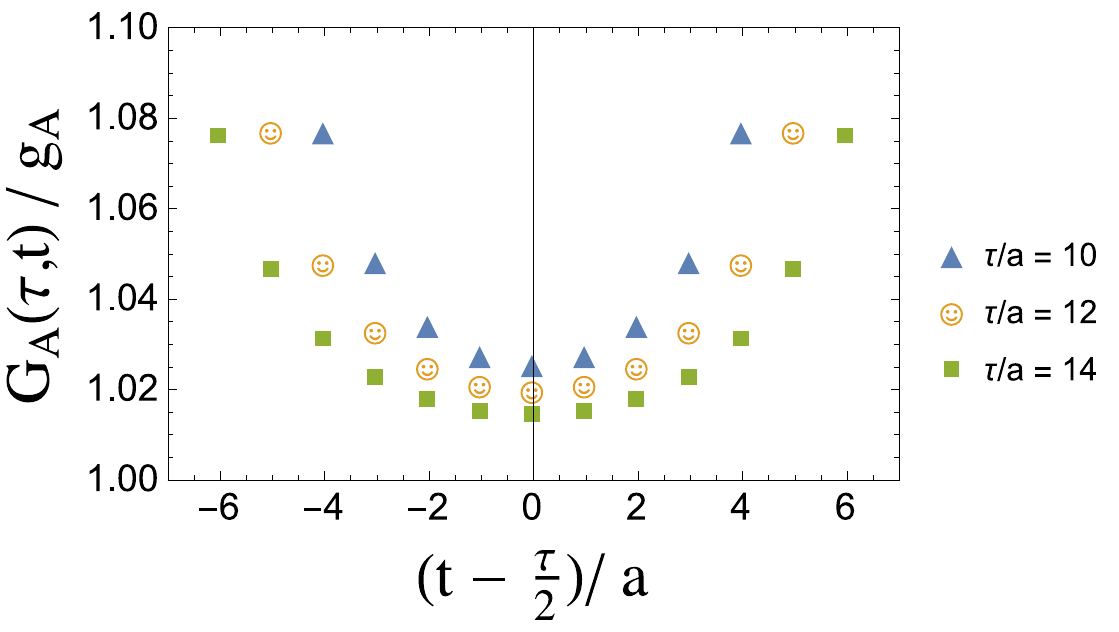}%
\caption{
Excited-state contamination in the axial-vector three-point function of the nucleon. 
The contamination function 
$G_A (\tau, t)$
is plotted as a function of the current insertion time 
$t$ 
(measured relative to the midpoint between source and sink), 
for three values of the source-sink separation, 
$\tau / a = 10$, 
$12$, 
and 
$14$. 
Excited states with pions tend to drive the axial correlator upwards, 
which could potentially lead to an overestimation of 
$g_A$
of a few percent at most.}
\label{f:GA}
\end{figure}
%
%
%

In Fig.~\ref{f:GA}, 
we plot the chiral perturbation theory evaluation of the three-point function ratio
$G_A(\tau,t)$. 
Notice that in the dual limit of 
$\tau \gg t$
and 
$t \gg 1$, 
we have 
$G_A (\tau, t) \to g_A$. 
What is of interest concerns how the three-point function ratio approaches the axial charge, 
i.e.~from above or below. 
Unlike the two-point function, 
the couplings
$g_{n,m}$
are real valued, and therefore not necessarily positive. 
Looking at Fig.~\ref{f:GA}, 
we see that pion-nucleon contamination drives the axial current correlation function upwards. 
In the plot, 
we show the axial current correlator ratio as a function of the current insertion time 
$t$
for three values of the source-sink separation, 
$\tau / a = 10$, $12$, and $14$, 
with 
$a = 0.1 \texttt{fm}$
taken as a typical value for the lattice spacing, 
on a lattice of size
$L = 48 a$. 
Because the pion-nucleon excited states drive the axial current correlator upwards, 
these contributions could lead to an overestimation of the axial charge if not properly accounted for. 
This trend is opposite lattice QCD data for the axial charge, 
which overwhelmingly approach a plateau from below. 
For a summary of lattice QCD results for the nucleon, 
see%
~\cite{Syritsyn:2014saa,Constantinou:2014tga}.

\section{Summary and Outlook}

Above we see how chiral perturbation theory can be used to address pion-nucleon and pion-delta contributions to nucleon correlation functions. 
The applicability of chiral perturbation theory for the lattice nucleon interpolating fields requires that the smearing radius of the lattice operator is considerably smaller than the Compton wavelength of the pion, 
$R_\mathcal{O} \ll m_\pi^{-1}$. 
In this regime, 
the lattice interpolating field can be systematically mapped into local operators in chiral perturbation theory. 
This mapping gives one the ability to compute excited-state contamination to lattice nucleon correlation functions. 
In particular, 
the perturbative pion-nucleon and pion-delta contamination.

In considering the nucleon two-point function in a general context, 
we find that phase-space arguments strongly disfavor large pion-nucleon contamination. 
In chiral perturbation theory, 
there is additional suppression due to the strength of pion-nucleon interactions, 
$\sim g_A^2 / ( f L)^2$. 
For weak pion-nucleon interactions, 
these results suggest that excited-state contributions to nucleon two-point functions would correspond to excited-state hadrons (in the case where they are bound states).
Consideration of the axial-vector current matrix elements in the nucleon, 
we see that the dominant excited-state contamination arises from 
nucleon to pion-nucleon intermediate states. 
For insufficient time separation between source, sink, and current insertion time, 
such excited-state contamination drives the axial correlation functions upward. 
This would lead to an overestimation of the nucleon axial charge, 
$g_A$, 
if such contributions are not properly removed. 
On the basis of the perturbative pion-nucleon interactions in chiral perturbation theory, 
we conclude that underestimation of the nucleon axial charge is not likely due to pion-nucleon excited-state contamination.

Looking to future work, 
it would be interesting to compute the excited-state contributions to other observables determined from lattice three-point functions, 
such as the nucleon scalar and tensor charges, quark momentum fraction in the nucleon, \emph{etc}. 
These results, 
at minimum, 
would provide an indication of the sign of excited-state contamination in these quantities, 
and this can be compared with the actual behavior of the lattice correlation functions. 
Furthermore, 
lattice QCD studies of correlation functions using well-localized nucleon interpolating fields would allow confrontation with these predictions from chiral perturbation theory. 
For the case of resonances, 
our work shows the necessity of including pion-nucleon operators. 
This is, 
by now, 
not at all surprising to lattice practitioners. 
The case of resonances is additionally interesting due to the non-perturbative nature of the pion-nucleon interactions. 
In this respect, 
it would be beneficial to understand nucleon, pion-nucleon, and resonance contributions to correlation functions in the context of unitarized chiral perturbation theory models.


\end{document}